# Reversing ferroelectric polarization in multiferroic $DyMn_2O_5$ by nonmagnetic Al substitution of Mn

Z. Y. Zhao, M. F. Liu, X. Li, J. X. Wang, Z. B. Yan, K. F. Wang, and J. –M. Liu[a]

*Laboratory of Solid State Microstructures, Nanjing University, Nanjing 210093, China*

[**Abstract**] The multiferroic $RMn_2O_5$ family, where R is rare-earth ion or Y, exhibits rich physics of multiferroicity which has not yet well understood, noting that multiferroicity is receiving attentions for promising application potentials. $DyMn_2O_5$ is a representative member of this family. The ferroelectric polarization in $DyMn_2O_5$ is claimed to have two anti-parallel components: one ($P_{DM}$) from the symmetric exchange striction between the $Dy^{3+}$-$Mn^{4+}$ interactions and the other ($P_{MM}$) from the symmetric exchange striction between the $Mn^{3+}$-$Mn^{4+}$ interactions. We investigate the evolutions of the two components upon a partial substitution of $Mn^{3+}$ by nonmagnetic $Al^{3+}$ in order to tailor the Mn-Mn interactions and then to modulate component $P_{MM}$ in $DyMn_{2-x/2}Al_{x/2}O_5$. It is revealed that the ferroelectric polarization can be successfully reversed by the Al-substitution via substantially suppressing the $Mn^{3+}$-$Mn^{4+}$ interactions and thus the $P_{MM}$. The $Dy^{3+}$-$Mn^{4+}$ interactions and the polarization component $P_{DM}$ can sustain against the substitution until a level as high as $x$=0.2. In addition, the independent Dy spin ordering is shifted remarkably down to an extremely low temperature due to the $Al^{3+}$ substitution. The present work not only confirms the existence of the two anti-parallel polarization components but also unveils the possibility of tailoring them independently.

---

[a] To whom all correspondences should be addressed, E-mail: liujm@nju.edu.cn

## I. Introduction

In recent years there has been an upsurge in research into multiferroic materials that display coupling between magnetic and ferroelectric orders [1, 2]. In particular, those multiferroics with magnetically induced electric polarization (so-called type-II multiferroics) have been receiving attention [2-4]. Rare-earth manganites $RMn_2O_5$ (RMO) represent a specific class of such materials featured with large electric polarization and complicated magnetic interactions [5-10], with respect to other well known multiferroics such as $RMnO_3$ [11-13], $LiCu_2O_2$ [14], and $MnWO_4$ [15] etc. Due to the complexities in lattice structure and magnetic interactions, our understanding of the microscopic mechanisms underlying the ferroelectricity and the complicated magnetic transitions of $RMn_2O_5$ is still in a stage of continuous updating, which needs substantial efforts [16-18].

It is noted that all the members of the $RMn_2O_5$ family have similar structural ingredients [10, 19]. The Mn ions are partitioned into $Mn^{3+}$ and $Mn^{4+}$, which are coordinated respectively in square pyramid Mn-O units and octahedral Mn-O units. The octahedra and pyramids are corner-sharing by either the pyramid base or pyramid apex, and the adjacent pyramids are connected with their bases. Along the *c*-axis, the octahedra sharing edges constitute linear chains. Each $Mn^{3+}$ ion is located in between two $Mn^{4+}$ ions, and the $R^{3+}$ ions are located on the alternative layer between two $Mn^{4+}$ ions. Therefore, $RMn_2O_5$ can be written as $R^{3+}Mn^{3+}Mn^{4+}5O^{2-}$. Obviously, the two structural blocks (pyramid Mn-O units and octahedral Mn-O units) stack alternatively and constitute different stacking sequences along the three orthogonal axes, making a number of degrees of freedom for structural distortions and magnetic interactions [16, 20]. The multifold competing interactions in $RMn_2O_5$ mainly arise from the multi-valance states of Mn and serious lattice distortions associated with the above mentioned structural ingredients, resulting in a set of complicated magnetic transitions. If the R ion has large magnetic moment, the magnetic transitions can be even more complicated due to the non-negligible R-Mn interactions, noting that the 4*f*-magnetism is quite different from the 3*d*-magnetism [21].

For a detailed illustration, we take $DyMn_2O_5$ as an example, and the *ab*-plane projected structural model is shown in Fig.1(a) where the Dy and Mn spin structures are schematically plotted, referred to the lattice and spin structures data from literature [22]. It is seen that the lattice accommodates the interactions from the $Mn^{3+}$-$Mn^{4+}$, $Mn^{4+}$-$Mn^{4+}$, $Mn^{3+}$-$Mn^{3+}$, $Dy^{3+}$-$Mn^{4+}$, $Dy^{3+}$-$Mn^{3+}$, and $Dy^{3+}$-$Dy^{3+}$ pairs. The paramagnetic phase above temperature $T\sim43K$ transits into an incommensurate antiferromagnetic (IC-AFM) phase, followed by a commensurate AFM (C-AFM) phase below $T_{N1}\sim40K$, and then by the coexistence of an IC-AFM phase and a C-AFM phase below $T_{N2}\sim28K$. This coexistence is again replaced by two coexisting IC-AFM phases below $T_{N3}\sim20K$. At $T<T_{Dy}\sim8K$, the $Dy^{3+}$ spins order independently [18]. Recently, the noncollinear Mn spin order with helical or cycloidal geometry in $DyMn_2O_5$ was reported [19]. The ferroelectric transitions associated with the magnetic transition sequence were investigated, and so far reported data are somehow controversial [7, 8]. Both the C-AFM and IC-AFM phases can be ferroelectric although the IC-AFM phase may not. Basically, the electric polarization $P$ most likely aligns along the $b$-axis, but its $T$-dependence appears to be complicated and no consistency is reached owing to the lacking of sufficient data [7, 8, 23].

Even though those unclear issues mentioned above are under investigations, a simplified scenario on the ferroelectricity can be outlined, regardless of the details of the magnetic transitions around $T_{N2}$ and $T_{N3}$. It is not strange that the ferroelectricity in $RMn_2O_5$ has complicated origins. In fact, earlier first-principles calculations predicted an electric polarization much larger than measured one. The reason is that the polarization has two anti-parallel components which originate respectively from the electronic and ionic contributions [24]. On the other hand, it was suggested that the Mn-Mn and Dy-Mn interactions both make contributions to the ferroelectricity via the symmetry exchange strictions, as schematically shown in Fig.1(c)-(d) [23]. The whole lattice structure can be mapped into a spatial pattern filled with two types of block groups alternatively. One is block A which consists of a $Mn^{4+}$-O octahedron sandwiched with two $Dy^{3+}$-O units, and the other is block B which consists of a $Mn^{4+}$-O octahedron sandwiched with two $Mn^{3+}$-O pyramids. The spins in the block A ($Dy^{3+}$-$Mn^{4+}$-$Dy^{3+}$) roughly align in the ↓↓↑ or ↑↑↓ pattern along the $b$-axis although there are small components along the $a$-axis and $c$-axis, generating a local electric dipole as indicated by the inner-open arrow $P_{DM}$ in Fig.1(c). The spin alignment in the block B ($Mn^{3+}$-$Mn^{4+}$-$Mn^{3+}$) takes roughly the ↓↑↑ or ↑↓↓ pattern along the $b$-axis too, as shown in Fig.1(d), generating a local electric dipole as indicated by the inner-open arrow $P_{MM}$.

The two types of electric dipoles are roughly anti-parallel to each other. It is noted that the two types of block groups stack alternatively and occupy the whole lattice. The macroscopic ferroelectric polarization $P$ has two anti-parallel components $P_{DM}$ and $P_{MM}$. Therefore, $DyMn_2O_5$ is basically a ferrielectric rather than a normal ferroelectric [7, 23].

The above model on the magnetically induced ferroelectricity may be still simplified and details of the correlations between the ferroelectricity and magnetic transitions around $T_{N2}$ and $T_{N3}$ are in fact not yet understood. We don't deal with these details in this work, and our major concern goes to the qualitative scenario with which the origin of $P_{DM}$ and $P_{MM}$ can be understood and the tuning possibility can be explored. Recently, the ferrielectricity of $DyMn_2O_5$ was discussed in the above model framework with the measured electric polarization $P=P_{DM}+P_{MM}$, and their $T$-dependences are schematically shown in Fig.2 [23]. Polarization $P$ is negative at $T<T_{N1}$ and becomes positive at low $T$, giving a sign reversal at certain point $T_{P=0}$. It is because of $P_{MM}>0$ and $P_{DM}<0$ so that their superposition ($P=P_{DM}+P_{MM}$) constitutes the complicated $T$-dependence and sign reversal at $T_{P=0}$. What motivates us here is how to modulate the electric polarization using some approaches, in particular the possibility of reversing/tailoring the polarization.

In attempt to tailor the electric polarization, however, the complexity is added by the independent Dy spin ordering below $T_{Dy}$ [7, 8, 23]. On one hand, the independent Dy spin ordering destabilizes significantly the ↓↓↑ and ↑↑↓-type $Dy^{3+}$-$Mn^{4+}$-$Dy^{3+}$ spin order in block A, thus suppressing partially the polarization component $P_{DM}$. This partial suppression is the reason for the anomalous slowing-down of the increasing tendency of $P$ with decreasing $T$ below $T_{Dy}$, as shown in Fig.2 for a guide of eyes, noting that this increasing tendency would be otherwise continuing, as shown by the $P'$ as a function of $T$. However, due to the weak $Dy^{3+}$-$Dy^{3+}$ interaction, the independent Dy spin ordering is sensitive to the structural and magnetic perturbations and can be easily destabilized. It is suggested that any tailoring approach, if effective in tailoring the two polarization components $P_{MM}$ and $P_{DM}$, may impose substantial impact on the independent Dy spin ordering as a side-effect.

Looking back at Fig.1, one sees that the $Mn^{4+}$-O octahedron is centered at the units of the two types of blocks. It is imagined that a substitution of $Mn^{4+}$ may disorder the spin alignments in both the block A and block B, thus damaging the $P_{MM}$ and $P_{DM}$ unfavorably for the ferroelectricity. If any approach to substitute partially $Dy^{3+}$ or $Mn^{3+}$ is taken, one of the two types of blocks would be tailored while the other may sustain without much change, as long as the substitution is limited at a low level. Surely, such a tailoring approach must rely on structural perturbations as small as possible, which is yet tough if not impossible. This is the main motivation for the present work.

We consider the $Al^{3+}$-substitution of Mn, since $Al^{3+}$ is nonmagnetic and its ionic radius is only slightly smaller than $Mn^{3+}$ ($Mn^{4+}$), so that the change of lattice environment remains small. We shall demonstrate a reversal of electric polarization in $DyMn_2O_3$ by gradually removing the ↓↑↑ or ↑↓↓ alignment in the block B, as tentatively shown in Fig.1(e), where component $P_{MM}$ disappears gradually but component $P_{DM}$ survives. Furthermore, this substitution also lowers the stability of the independent Dy spin order below $T_{Dy}$, making the component $P_{DM}$ to follow the $T$-dependence of $P'_{DM}$ shown in Fig.2. The present approach obviously allows a reversing of the ferroelectric polarization while the substitution strategy is so simple.

In this work, we prepare a series of polycrystalline $DyMn_{2-x/2}Al_{x/2}O_5$ samples. It will be shown that the $Al^{3+}$ ions most likely replace $Mn^{3+}$ ions rather than $Mn^{4+}$ ions, as revealed by various structural and chemical characterizations. The measured data on the electric and magnetic properties confirm a successful reversal of the ferroelectric polarization by partially substituting $Mn^{3+}$ with $Al^{3+}$. Furthermore, several additional multiferroic behaviors are observed, suggesting rich physics in terms of ferroelectric response to the tailoring of magnetic orders.

## II. Experimental details

### A. Samples preparation & structural characterization

A series of polycrystalline $DyMn_{2-x/2}Al_{x/2}O_5$ samples were prepared by standard solid state sintering. Stoichiometric amount of $Dy_2O_3$ (99.99%), $Mn_2O_3$ (99%), and $Al_2O_3$ (99.99%) was thoroughly mixed (ground), compressed into pellets, and sintered at 1200°C for 24 h in a flowing oxygen atmosphere with several cycles of intermediate grindings. For every sintering cycle, the samples were cooled down to room temperature at a rate of 100°C per hour from the sintering temperature. The as-prepared samples were cut into various shapes for subsequent microstructural and property characterizations. For structural characterization, the crystallinity was checked using X-ray diffraction (XRD) with Cu $K\alpha$ radiation at room temperature and the high-resolution data were used for structural Rietveld refining by the GSAS program. The atomic ratios and chemical valence states of Mn species were probed using the X-ray photoelectron spectroscopy (XPS, PHI500 Versa Prove, UIVAC-PHI Inc.) with the monochromatic Al $K\alpha$ radiation. Particular attention was paid to the relative variations of the $Mn^{3+}/Mn^{4+}$ ions so that the Al-substitution of $Mn^{3+}$ site can be checked.

*B. Measurements of magnetic and electric properties*

The specific heat ($C_P$), magnetization ($M$) and *dc* magnetic susceptibility ($\chi$), dielectric constant ($\varepsilon$) and electric polarization ($P$) of the samples were characterized. The $M$ and $\chi$ were measured using the Quantum Design Superconducting Quantum Interference Device (SQUID) in the zero-field cooled (ZFC) mode and field-cooling (FC) mode, respectively. Both the cooling field and measuring field are 1.0kOe. The $C_p$ was measured using the Quantum Design Physical Properties Measurement System (PPMS) in the standard procedure.

Here the measurement of pyroelectric current $I_{pyro}(T)$ was critical since the ferroelectric polarization $P(T)$ was evaluated from the $I_{pyro}(T)$ data. Each sample was polished into a thin disk of 0.2mm in thickness and 10mm in diameter, and then sandwich-coated with Au layers as top and bottom electrodes. The measurement was performed using the Keithley 6514A and 6517 electrometers connected to the PPMS and the related details were described in Ref. [23]. The pre-poling electric field $E_{pole}$ was 10kV/cm and the samples were cooled under the electric poling down to $T_{end}$=2K. It is noted that the as-prepared polycrystalline samples are highly insulating and the recorded background current noise amplitude was less than 0.2pA. The polarization $P(T)$ was obtained by integrating the collected $I_{pyro}(T)$ data, from $T_0 >> T_{N1}$ down to $T_{end}$.

The validity of this procedure was confirmed repeatedly in earlier works and here it is confirmed again. Fig.3 shows the measured $I_{pyro}(T)$ data at three warming rates (2, 4, 6K/min) for sample *x*=0.0. It comes to our attention that the three curves, if normalized by the corresponding warming rate, almost perfectly overlap with each other, showing no difference between them within the measuring uncertainties and less than 0.3K peak-to-peak shift along the *T*-axis. These peaks are sharp and well fixed while thermally stimulated currents other than the pyroelectric current are usually broad. These features indicate that the measured data do come from the pyroelectric current without identifiable contribution from other sources. In addition, the measured $I_{pyro}$-*T* curve can be switched upon a reversed poling field.

In addition, the *ε*(*T*) data at various frequencies were collected with an *ac*-bias of ~50mV. Besides the *ε*(*T*) data and *P*(*T*) data, we also measured the response of *P* to magnetic field *H* in two modes. One is the isothermal mode with which the variation in *P* in response to the scanning of *H* was detected and the other is the iso-field mode with which the *P-T* data under a fixed *H* were collected. By such measurements, one can evaluate the ME coupling by defining Δ*P*(*H*)=*P*(*H*)-*P*(*H*=0) as the magnetoelectric (ME) parameter.

### III. Results and discussion

*A. Structural distortion and chemical valence states*

We first check the crystallinity and associated lattice distortion. The XRD *θ*-2*θ* spectra for several samples are plotted in Fig.4(a). While all these reflections can be indexed by the standard database of orthorhombic $DyMn_2O_5$ lattice, and gradual peak shifts towards the high-angle side are observed. It is noted that the Al-substitution of Mn up to *x*=0.20 does not change the lattice symmetry. The inset in Fig.4(b) shows the local peaks at 2*θ*~29° for *x*=0.00, 0.10, and 0.16, and the local peak profiles are almost the same although the profile shifts gradually rightward with increasing *x*. We perform the Rietveld refining of the data and one example is given for sample *x*=0.04, as shown in Fig.4(b). The refining reliability is as high as $R_{wp}$=4.48%. The evaluated lattice unit volume *V* as a function of *x*, plotted in Fig.3(c), decreases gradually, due to the slightly smaller ionic size of $Al^{3+}$ than Mn ion, and the fitted errors are reasonably small. What should be mentioned here is that the *V*(*x*) seems to be slightly nonlinear rather than linear as predicted by the Vegard's law, suggesting that the possible cation charge valence variation upon the Al-substitution can't be excluded although this variation may be ascribed to the variation of magnetic interactions via the spin-lattice coupling. Therefore, careful checking of the charge valence of Mn and Al should be made.

We consult to the XPS determination of Mn ions in valence state upon the Al-substitution. The data on all these samples show no trace of intensity from the $Mn^{5+}$ or $Mn^{2+}$ within the apparatus resolution. Since the binding energies for the $Mn^{3+}2p_{1/2}/Mn^{4+}2p_{1/2}$ and $Mn^{3+}2p_{3/2}/Mn^{4+}2p_{3/2}$ are very close respectively [25], a highly reliable fitting of the XPS peaks associated with Mn is required. Fig.5(a) show this fitting for sample $x$=0.0, in which the $2p_{1/2}$ and $2p_{3/2}$ peaks are assumed to be the superimposition of the contributions from the $Mn^{3+}$ and $Mn^{4+}$. It is seen that the sum of the shadow areas below the peaks $Mn^{3+}2p_{1/3}$ and $Mn^{3+}2p_{3/2}$ is roughly equal to that below the peaks $Mn^{4+}2p_{1/3}$ and $Mn^{4+}2p_{3/2}$, indicating that the $Mn^{3+}$:$Mn^{4+}$ ratio is close to the nominal value. The same fitting procedure is applied to the other samples and the fitted $Mn^{3+}$:$Mn^{4+}$ ratios are consistent with the nominal ones with an uncertainty of ±5%.

Alternatively, we present in Fig.5(b) the amplified Mn $2p_{3/2}$ peaks for several samples to see the overall tendency of the peak shift. Indeed no trace signals from $Mn^{5+}$ or $Mn^{2+}$ are seen. The gradual shifting of the peak towards the high-energy side with increasing $x$ is concededly identified, indicating more $Mn^{4+}$ ions than $Mn^{3+}$ ions in the samples with higher $x$. This behavior provides direct evidence on the claim that the $Al^{3+}$ ions substitute the $Mn^{3+}$ ions rather than $Mn^{4+}$ ions. Although one can't exclude the possibility for tiny occupation of the $Mn^{4+}$ sites by $Al^{3+}$, the present XPS data demonstrate that the $Al^{3+}$-occupation of $Mn^{3+}$ is dominant, favored from the point of view of charge balance.

*B. Anomalies of multiferroic properties at the magnetic phase transitions*

Before proceeding to the Al-substitution effects, we first look at the anomalies of several parameters at the magnetic phase transitions. The normalized specific heat $C_P/T$, magnetization $M$ (under ZFC and FC modes both), dielectric permeability $\varepsilon$, pyroelectric current $I_{pyro}$ ($I_{tot}$), and electric polarization $P$, are plotted in Fig.6. For a reference, the ferroelectric phases in various $T$-ranges are marked on the top row, including the recently confirmed ferroelectric X-phase [8, 23]. Parameters $C_P/T$, $\varepsilon$, $I_{pyro}$, and $P$, all show clear anomalies at the magnetic phase transition points $T_{N1}$, $T_{N2}$, $T_{N3}$, and $T_{Dy}$. However, the $M(T)$ curves are trivial except the broad peak at $T_{Dy}$. This behavior is well known and the reason is that the paramagnetic fluctuations from the Dy spins above $T_{Dy}$ are dominant, submerging the anomalies from the Mn spin ordering. The anomalies of $\varepsilon$, $I_{pyro}$, and $P$ at these transition points reflect the ME coupling. In particular, the anomalies of the $I_{pyro}(T)$ curve at $T_{N1}$, $T_{N2}$, $T_{N3}$, and $T_{Dy}$, as shown in Fig.6(d), are one-to-one corresponding to those in the $C_P/T(T)$ curves but more remarkable than the latters.

For sample $x$=0.0, earlier experiments established the correlations between these peaks in the $I_{pyro}(T)$ curve, the ferroelectric transitions associated with $P_{MM}$ and $P_{DM}$, and the magnetic phase transitions [23]. A brief description is given here as an additional illustration to Fig.2. First, the sharp negative current peak right below $T_{N1}$ indicates the generation of $P_{MM}$ (<0) due to the development of roughly collinear $Mn^{3+}$-$Mn^{4+}$-$Mn^{3+}$ spin order and the $P_{MM}$ tends to be saturated at $T \sim T_{N2}$ and below. Second, the broad positive current bump around $T_{N2}$ seems to sign the generation of $P_{DM}$ (>0) due to the development of the $Dy^{3+}$-$Mn^{4+}$-$Dy^{3+}$ spin order and the $P_{DM}$ increases continuously with decreasing $T$. Third, a Dy-Mn spin coupling occurs which can be understood as the Dy spin ordering induced by the Mn spin orders. This induced Dy spin ordering may initiate above $T_{N2}$ but develop well below $T_{N3}$. However, details of this coupling remain unclear. Fourth, the sharp positive peak around $T_{Dy}$ signs the consequence of the independent Dy spin ordering, which can damage the collinear $Dy^{3+}$-$Mn^{4+}$-$Dy^{3+}$ spin order and thus the $P_{DM}$, while the $P_{MM}$ is less affected. What should be answered is the question why the $I_{pyro}(T)$ peak right below $T_{N1}$ is negative, i.e. $P_{MM}$<0. The reason is that $P_{DM}$>$P_{MM}$ at $T_{end} \sim$2K and therefore the cooling under electric poling down to $T_{end}$ enables the $P_{DM}$ to align along the poling field but $P_{MM}$ is antiparallel to the field, i.e. $P_{MM}$<0 [23]. In Fig.6(e), the $P_{MM}$ and $P_{DM}$ as a function of $T$ respectively are plotted and $P=P_{MM}+P_{DM}$. In the following, one will see that these correlations are the basis on which the effects of Al-substitution are understood.

We first investigate the effect of Al-substitution on the independent Dy spin ordering at $T_{Dy}$. The $M$-$T$ data at several $x$ values are plotted in Fig.7. It is seen that the $T_{Dy}$ is indeed suppressed with increasing $x$. For $x$=0.20, $T_{Dy}$ falls down to ~2K and below. At the first glance this phenomenon seems unusual and our evidence supports that the Al ions substitute the Mn ions rather than the Dy ions. Nevertheless, due to the high sensitivity of the independent Dy spin ordering to weak perturbations of lattice and spin interactions [21], the Al-substitution seems to generate such perturbations sufficient for disordering this order. This implies that the Al-substitution most likely removes the influence of the independent Dy spin ordering on the ferroelectricity (mainly component $P_{DM}$), enabling the physics simpler. The tendency of $P_{DM} \rightarrow P'_{DM}$ and $P \rightarrow P'$, as predicted in Fig.2, will be confirmed below.

### C. Ferroelectric polarizations

Now we focus on the electric polarization in response to the Al-substitution. The $P$-$T$ data for a series of samples are plotted in Fig.8(a)-(g). We also insert the $I_{pyro}(T)$ data for sample $x$=0.0 in each plot for a comparison so that the change of the $I_{pyro}(T)$ curve with increasing $x$ can be identified easily.

Several interesting characters of the $I_{pyro}(T)$ curve and $P(T)$ curve with increasing $x$ are worthy of addressing. First, as expected, the negative current peak right below $T_{N1}$ is remarkably suppressed and downshifted with increasing $x$, indicating the suppression of the ferroelectric transitions associated with the $P_{MM}$. At $x\sim0.08$ and above, the negative current peak becomes nearly disappeared. Correspondingly, component $P_{MM}$ becomes disappear. It is suggested that the Al-substitution does succeed in breaking the $Mn^{3+}$-$Mn^{4+}$-$Mn^{3+}$ collinear spin order. Second, the positive current peak around $T_{Dy}$ in sample $x$=0.0, as indicated by the arrow in Fig.8(a), becomes very weak in sample $x$=0.02 and disappeared in sample $x$=0.04 and other samples with higher $x$. Referring to the magnetic data shown in Fig.7, it is suggested that the peak disappearance at $x\geq0.04$ is attributed to the Al-substitution induced downshifting of the independent Dy spin ordering. This feature is consistent with the predicted $P'$ as a function of $T$ in Fig.2. Third, the broad current bump around $T_{N3}$ in sample $x$=0.0 is evolved into a sharp peak locating in-between $T_{N2}$ and $T_{N3}$ for low $x$ level ($x$<0.08). It is found that, upon increasing $x$ from 0.0 to 0.08, this positive peak increases in height and shifts rightward, in compensation with the height decreasing and leftward shifting of the negative peak around $T_{N1}$. Eventually, the two peaks meet and annihilate with each other. The last 'moment' of the two peak annihilation is shown in Fig.8(d) at $x$=0.08. One easily understands that the two peaks correspond respectively to the polarization generation at the higher-$T$ side (negative current peak) and polarization disappearance at the lower-$T$ side (positive current peak). The simultaneous evolution of the two peaks illustrates how the $P_{MM}$ is suppressed upon the Al-substitution, as shown in Fig.8(a)~(d).

In accompanying with the serious suppression of the $P_{MM}$, the evolution of the $P_{DM}$ is different, although a quantitative evaluation is impossible since we only have the $P(T)$ data.

Without doubt, the Al-substitution certainly damages the $P_{DM}$ too because the spin structure is anyhow diluted by the $Al^{3+}$ ions. This effect can be seen in Fig.8, and the measured $P$ in the low-$T$ range does not increase much while the negative $P_{MM}$ is seriously suppressed by the Al-substitution. Since $P=P_{DM}+P_{MM}$, it is clear that the $P_{DM}(T)$ decreases with increasing $x$. As $x \geq 0.08$, only the $P_{DM}$ is left while the $P_{MM}$ is completely suppressed. However, the $P_{DM}$ is much more robust than the $P_{MM}$ against the Al-substitution and the $P_{DM}>100\mu C/m^2$ at 2K is retained even $x=0.20$. This implies that the Al-substitution does not change much the collinear $Dy^{3+}$-$Mn^{4+}$-$Dy^{3+}$ spin order. To show this, the as-evaluated maximal $P_{DM}$ and $P_{MM}$ values are plotted as a function of $x$ in Fig.9. Unfortunately, further increasing of $x$ generates impurity phase and the stability of this spin order over the whole $x$-range may be concerned by other approaches.

*D. Polarization reversal*

It is interested to note that the significant impact of the Al-substitution on $P_{MM}$ as described above leads to a negative-positive reversal of polarization $P$, which is illustrated by plotting the $P(x)$ curves at several $T$, as shown in Fig.8(h). The $P(T=2K)$ data are always positive due to $P_{DM}>P_{MM}$, but both the $P(T=15K)$ and $P(T\sim T_{N3})$ reverse their signs from negative to positive at certain $x$ ($x\sim0.04$ and 0.065). The successful reversal can be more clearly seen by the phase-diagram in Fig.10. The whole phase-diagram is divided into three regions. Given $P_{MM}<0$ and $P_{DM}>0$, one has $P=(P_{MM}+P_{DM})<0$ while $|P_{MM}|>>|P_{DM}|$ in region I. In region II, $|P_{MM}|>|P_{DM}|$ is replaced by $|P_{MM}|<|P_{DM}|$ so we have $P=(P_{MM}+P_{DM})>0$. A crossing through the boundary between regions I and II is accompanied with a reversal of polarization $P$. One has $P_{MM}\sim0$ so $P>0$ in region III, while $P_{DM}\sim0$ and $P_{MM}\sim0$ are expected as $x>0.20$. The coarse solid double-head arrow indicates a reversal of polarization $P$ while the dashed double-head arrow shows the generation/disappearance of component $P_{MM}$ associated with the Mn-Mn interactions.

One may address that such a polarization reversal in multiferroics with magnetically induced ferroelectricity has rarely been observed so far and this is the first experimental evidence for the polarization reversal by chemical substitution. While the complexity of the multiferroic physics in $DyMn_2O_5$ is well known [7, 8], here we present a simple example for this physics, which allows a direct manipulation of the ferroelectric polarization in $RMn_2O_5$.

### E. Magnetoelectric response

Keeping in mind the motivations and consequences of the Al-substitution in $DyMn_2O_5$, additional evidence with the proposed physics therein can be obtained from the magnetoelectric response. As noted earlier, the $Dy^{3+}$-$Dy^{3+}$ spin interaction is quite weak and thus a remarkable response of the $Dy^{3+}$ spin order to magnetic field is expected. Differently, the $Mn^{3+}/Mn^{4+}$ spins are much more robust [8, 23]. This argument can be checked here since $P_{MM}$ is nearly zero in region III (Fig.10), by measuring the magnetoelectric response. We present in Fig.11 the $P(T)$ data obtained at several $H$ in the iso-field mode, for three samples $x$=0.0, 0.12, and 0.16.

One can easily understand the observed results in Fig.11. For sample $x$=0.0, the consequence of applying a magnetic field should be the rapid suppression of $P$ from positive to negative in the low-$T$ range while no big variation of $P$ in the high-$T$ range [23]. The data in the left column do confirm this consequence, and the $P$ value at $T$=2K falls from ~120μC/m$^2$ under $H$=0 to -95μC/m$^2$ under $H$=5T, noting that the anomalous slowing-down effect of $P$ below $T_{Dy}$ becomes disappeared as $H$>1.0T. For sample $x$=0.12, noting no more $P_{MM}$ available, the measured $P$ initiates roughly at $T_{N3}$<$T$<$T_{N2}$ and then gradually increases with decreasing $T$, indicating the ferroelectric phase transitions associated with $P_{DM}$, as shown in the middle column of Fig.11. No anomaly of the $P(T)$ dependence below $T_{Dy}$ is detected, obviously due to the absence of the independent Dy spin ordering. The electric polarization is remarkably suppressed upon implication of magnetic field, by a change of ~50% at $H$=5.0T. It is noted that the ferroelectric transition point does not change much. Similar behaviors are identified for sample $x$=0.16 and the magnetoelectric response is even more remarkable, as seen in the right column of Fig.11. For both the latter two cases, the measured $P$ remains positive under a field as high as $H$=9.0T, very different from the case observed for sample $x$=0.0.

The complete disappearance of the electric polarization is expected when the substitution is higher than $x$=0.20, which is not available unfortunately for us at the current synthesis conditions. On the other hand, extremely high magnetic field is expected to suppress completely the polarization. Nevertheless, the so far available data are sufficient to confirm the proposed model (Fig.1 and Fig.2) and the phase diagram Fig.10, while the non-magnetic Al-substitution of Mn demonstrates its capability to reverse the electric polarization.

*F. Discussion*

It is highly agreed that $DyMn_2O_5$ and other $RMn_2O_5$ family members are complicated in terms of magnetic structures and ferroelectricity origins [16, 18]. However, the proposed model and the simple substitution strategy in this work enable our understanding of the physics of multiferroicity in a quite simplified framework. This framework seems to catch up the core of the physics, although the details of those relatively weak anomalies of magnetic and dielectric responses around $T_{N2}$ and $T_{N3}$ are not considered.

This simple strategy is essentially associated with the magnetic structure highlighted in Fig.1, noting that similar magnetic structure in $RMn_2O_5$ members with R=Gd, Tb, Ho, and Er, was reported recently [26-30]. In this sense, the present model would be of generality to some extent. Indeed, this model can be used to explain quite a number of multiferroic behaviors in these materials, bur many of them can't yet be reasonably predicted and interpreted. The possible reasons for these failures include the following aspects. First, the R ionic size variation is critical for the lattice distortion and thus the delicate balance of the multifold interactions. Second, the *4f-3d* coupling between R and Mn ions can be very different for different members. The Gd-Mn, Dy-Mn, and Ho-Mn couplings are strong while the Tb-Mn and Er-Mn couplings are relatively weak [26, 31]. For the latter cases, the polarization component *P* from the R-Mn coupling could be quite small and thus the underlying physics becomes different. In this sense, the lattice would accommodate a normal ferroelectric behavior rather than the ferrielectricity. Third, clear noncollinear spin components in these materials, as shown in Fig.1 for $DyMn_2O_5$, may contribute to the electric polarization via the spin-orbit coupling mechanism [20], which is not taken into account in the present work. This contribution, if available, would make the ferroelectric phase transitions and response of P to *T* and *H* different from the collinear three-spin block mechanisms addressed here. For example, the low-field response of *P* (Fig.11) may be more favorably contributed by this noncollinear mechanism. These issues deserve for further investigations, which, however, on the contrary suggests the substantial significance of the present experiments based on such a simple scenario.

## IV. Conclusion

In conclusion, we have investigated in details the effects of $Al^{3+}$ substitution of Mn ions on the magnetic and ferroelectric behaviors in multiferroic $DyMn_2O_5$, based on the proposed model for ferrielectricity generation. It is revealed that structurally the $Al^{3+}$ substitution favors the $Mn^{3+}$ sites rather than $Mn^{4+}$ ions, and makes the lattice contracting slightly. This tiny structural distortion seems to suppress remarkably the independent Dy spin ordering which enters below $T_{Dy}$~8K in $DyMn_2O_5$. In consequence, the ferrielectric lattice decomposes gradually into a normal ferroelectric lattice by disappearance of the polarization component $P_{DM}$, due to the gradual disordering of the ↓↓↑ or ↑↑↓ collinear $Mn^{3+}(Al^{3+})$-$Mn^{4+}$-$Mn^{3+}(Al^{3+})$ spin blocks, while the ↓↓↑ or ↑↑↓ collinear $Dy^{3+}$-$Mn^{4+}$-$Dy^{3+}$ spin blocks are maintained in the lattice. It is demonstrated that the simple strategy of the Al-substitution of Mn can be an effective approach to tune the electric polarization and reverse it from negative value to positive one. The present work provides a comprehensive understanding of the multiferroicity in $DyMn_2O_5$ and may shed light on efficient approaches to be taken for improving the multiferroic performances of the whole $RMn_2O_5$ systems.

**Acknowledgement**:

This work was supported by the National 973 Projects of China (Grants No. 2011CB922101), the Natural Science Foundation of China (Grants Nos. 11234005, 11374147, and 51332006), and the Priority Academic Program Development of Jiangsu Higher Education Institutions, China.

*References*:

1.  S. W. Cheong and M. Mostovoy, *Multiferroics: a magnectic twist for ferroelectrcity*, Nature Mater. 6, 13 (2007).

2.  G. Q. Zhang, S. Dong, Z. B. Yan, Y. Y. Guo, Q. F. Zhang, S. Yunoki, E. Dagotto, and J. –M. Liu, *Multiferroic properties of $CaMn_7O_{12}$*, Phys. Rev. B 84, 174413 (2011).

3.  T. Kimura, T. Goto, H. Shintani, K. Ishizaka, T. Arima, and Y. Tokura, *Magnetic control of ferroelectric polarization*, Nature (London) 426, 55 (2003).

4.  D. Khomskii, *Trends: classifying multiferroics: mechanisms and effects*, Physics 2, 20 (2009).

5.  N. Hur, S. Park, P. A. Sharma, J. S. Ahn, S. Guha, and S. W. Cheong, *Electric polarization reversal and memory in a multiferroic material induced by magnetic fields*, Nature (London) 429, 392 (2004).

6.  M. Fukunaga, Y. Sakamoto, H. Kimura, Y. Noda, N. Abe, K. Taniguchi, T. Arima, S. Wakimoto, M. Takeda, K. Kakurai, and K. Kohn, *Magnetic-field-induced polarization flop in multiferroic $TmMn_2O_5$*, Phys. Rev. Lett. 103, 077204 (2009).

7.  N. Hur, S. Park, P. A. Sharma, S. Guha, and S. W. Cheong, *Colossal magnetodielectric effects in $DyMn_2O_5$*, Phys. Rev. Lett. 93, 107207 (2004).

8.  D. Higashiyama, S. Miyasaka, N. Kida, T. Arima, and Y. Tokura, *Control of the ferroelectric properties of $DyMn_2O_5$ by magnetic fields*, Phys. Rev. B 70, 174405 (2004).

9.  K. F. Wang, J. –M. Liu, and Z. F. Ren, *Multiferroicity: the coupling between magnetic and ferroelectric orders*, Adv. Phys. 58, 321 (2009).

10. Y. Noda, H. Kimura, M. Fukunaga, S. Kobayashi, I. Kagomiya, and K Kohn, *Magnetic and ferroelectric properties of multiferroic $RMn_2O_5$*, J. Phys. Condens. Matt. 20, 434206 (2008).

11. T. Goto, T. Kimura, G. Lawes, A. P. Ramirez, and Y. Tokura, *Ferroelectricity and giant magnetocapacitance in perovskite rare-earth manganites*, Phys. Rev. Lett. 92, 257201 (2004).

12. O. Prokhnenko, R. Feyerherm, E. Dudzik, S. Landsgesell, N. Aliouane, L. C. Chapon, and D. N. Argyriou, *Enhanced ferroelectric polarization by induced Dy spin order in*

*Multiferroic DyMnO₃*, Phys. Rev. Lett. 98, 057206 (2007).

13. B. Lorenz, Y. Q. Wang, and C. W. Chu, *Ferroelectricity in perovskite HoMnO₃ and YMnO₃*, Phys. Rev. B 76, 104405 (2007).

14. S. Park, Y. J. Choi, C. L. Zhang, and S. W. Cheong, *Ferroelectricity in an S=1/2 chain cuprate*, Phys. Rev. Lett. 98, 057601 (2007).

15. K. Taniguchi, N. Abe, H. Sagayama, S. Ohtani, T. Takenobu, Y. Iwasa, and T. Arima, *Magnetic-field dependence of the ferroelectric polarization and spin-lattice coupling in multiferroic MnWO₄*, Phys. Rev. B 77, 064408 (2008).

16. A. B. Sushkov, M. Mostovoy, R. V. Aguilar, S. W. Cheong, and H. D. Drew, *Electromagnons in multiferroic RMn₂O₅ compounds and their microscopic origin*, J. Phys. Condens. Matter 20, 434210 (2008).

17. J. H. Kim, S. H. Lee, S. I. Park, M. Kenzenlmann, A. B. Harris, J. Schefer, J. H. Chung, C. F. Majkrzak, M. Takeda, S. Wakmoto, S. Y. Park, S. W. Cheong, M. Matsuda, H. Kimura, Y. Noda, and K. Kakurai, *Spiral spin structures and origin of the magnetoelectric coupling in YMn₂O₅*, Phys. Rev. B 78, 245115 (2008).

18. W. Ratcliff II, V. Kiryukhin, M. Kenzelmann, S.-H. Lee, R. Erwin, J. Schefer, N. Hur, S. Park, and S. W. Cheong, *Magnetic phase diagram of the colossal magnetoelectric DyMn₂O₅*, Phys. Rev. B 72, 060407(R) (2005).

19. P. G. Radaelli and L. C. Chapon, *A neutron diffraction study of RMn₂O₅ multiferroics*, J. Phys. Condens. Matter 20, 434213 (2008).

20. G. R. Blake, L. C. Chapon, R. G. Radaelli, S. Park, N. Hur, and S. W. Cheong, *Spin structure and magnetic frustration in multiferroic RMn₂O₅ (R=Tb, Ho, Dy)*, Phys. Rev. B 71, 214402 (2005).

21. R. A. Ewing, A. T. Boothryod, D. F. McMorrow, D. Mannix, H. C. Walker, and B. M. R. Wanklyn, *X-ray resonant diffraction study of multiferroic DyMn₂O₅*, Phys. Rev. B 77, 104415 (2008).

22. G. E. Johnstone, R. A. Ewing, R. D. Johnson, C. Mazzoli, H. C. Walker, and A. T. Boothroyd, *Magnetic structure of DyMn₂O₅ determined by resonant x-ray scattering*, Phys. Rev. B 85, 224403 (2012).

23. Z. Y. Zhao, M. F. Li, X.Li, L. Lin, Z. B. Yan, S. Dong, and J.-M. Liu, *Experimental

*observation of ferrielectricity in multiferroic DyMn$_2$O$_5$*, Sci. Rep. 4, 3984. (2014).

24. G. Giovannetti and J. van den Brink, *Electronic correlations decimate the ferroelectric polarization of multiferroic HoMn$_2$O$_5$*, Phys. Rev. Lett. 100, 227603 (2008).

25. T. Taniguchi, S. Mizusaki, N. Okada, Y. Nagata, S. H. Lai, M. D. Lan, N. Hiraoka, M. Itou, Y. Sakurai, T. C. Ozawa, Y. Noro, and H. Samata, *Crystallographic and magnetic properties of the mixed-valence oxides CaRu$_{1-x}$Mn$_x$O$_3$*, Phys. Rev. B 77, 014406 (2008).

26. N. Lee, C. Vecchini, Y. J. Choi, L. C. Chapon, A. Bombardi, P. G. Radaelli, and S. W. Cheong, *Giant tunability of ferroelectric polarization in GdMn$_2$O$_5$*, Phys. Rev. Lett. 110, 137203 (2013).

27. L. C. Chapon, G. R. Blake, M. J. Gutmann, S. Park, N. Hur, P. G. Radaelli, and S. W. Cheong, *Structural anomalies and multiferroic behavior in magnetically frustrated TbMn$_2$O$_5$*, Phys. Rev. Lett. 93, 177402 (2004).

28. C. Vecchini, L. C. Chapon, P. J. Brown, T. Chatterji, S. Park, S. W. Cheong, and P. G. Radaelli, *Commensurate magnetic structures of RMn$_2$O$_5$ (R=Y, Ho, Bi) determined by single-crystal neutron diffraction*, Phys. Rev. B 77, 134434 (2008).

29. U. Staub, Y. Bodenthin, M. García-Fernández, R. A. de Souza, M. Garganourakis, E. I. Golovenchits, V. A. Sanina, and S. G. Lushnikov, *Magnetic order of multiferroic ErMn$_2$O$_5$ studied by resonant soft x-ray Bragg diffraction*, Phys. Rev. B 81, 144401 (2010).

30. C. J. Wang, G. C. Guo, and L. X. He, *Ferroelectricity driven by the noncentrosymmetric magnetic ordering in multiferroic TbMn$_2$O$_5$: a first-principles study*, Phys. Rev. Lett. 99, 177202 (2007).

31. R. D. Johnson, C. Mazzoli, S. R. Bland, C-H. Du, and P. D. Hatton, *Magnetically induced electric polarization reversal in multiferroic TbMn$_2$O$_5$: Terbium spin reorientation studied by resonant x-ray diffraction*, Phys. Rev. B 83, 054438 (2011).

**Figure Captions**

Fig.1. A schematic drawing of the spin structure at low T and possible mechanisms for electric polarization generation. (a) The spin structure projected on the *ab* plane, and these structural units are not on the same lattice plane and they alternatively shift roughly 1/2 or -1/2 atomic layer distance along the *c*-axis. (b) The structural unit symbols used in (a). (c) The two sub-types of block A, and the local electric dipoles $P_{DM}$ due to the symmetric exchange striction associated with the ↓↓↑ and ↑↑↓ $Dy^{3+}$-$Mn^{4+}$-$Dy^{3+}$ spin alignments, respectively. (d) The two sub-types of block B, and the local electric dipoles $P_{MM}$ due to the symmetric exchange striction associated with the ↓↑↑ and ↑↓↓ $Mn^{3+}$-$Mn^{4+}$-$Mn^{3+}$ spin alignments, respectively. (e) The two sub-types of block C with the Al-substituted $Mn^{3+}$-$Mn^{4+}$-$Al^{3+}$ chains which have no electric dipole. The lattice of $DyMn_2O_5$ thus accommodates a ferrielectricity.

Fig.2. The predicted *T*-dependence of electric polarization *P* as a feature of ferrielectricity, where polarization *P* consists of two anti-parallel components $P_{DM}$ and $P_{MM}$. It is suggested that these polarizations should take the dependences $P'$, $P'_{DM}$ and $P_{DM}$ if no independent Dy spin ordering would occur below $T_{Dy}$.

Fig.3. Measured pyroelectric current $I_{pyro}(T)$ for DyMn2O5 with warming rate of 2K/min (a), 4K/min (a), and 6K/min (a).

Fig.4. (a) Measured XRD $\theta$-$2\theta$ spectra for a series of $DyMn_{2-x}Al_xO_5$ samples with labeled *x* values. (b) The Rietveld refined data for sample *x*=0.04 and the insert shows the local reflections of three samples *x*=0.00, 0.10, and 0.16. (c) The fitted unit volume V as a function of *x* with a fitting curve for guide of eyes.

Fig.5. (a) Measured XPS spectrum for sample *x*=0.0 as an example, where the $Mn^{3+}$ and $Mn^{4+}$ core energy levels are labeled and the measured peaks are decomposed as the superimposition of the contributions from the $Mn^{3+}$ and $Mn^{4+}$ excitations, respectively. (b) The local peaks around 642 eV corresponding to the Mn ($2p_{3/2}$) core level. The peak shifting toward the high

energy with increasing $x$ is shown, indicating the increasing $Mn^{4+}/Mn^{3+}$ ratio.

Fig.6. Measured specific heat $C_P/T$ (a), magnetization $M$ (b), dielectric constant $\varepsilon$ (c), pyroelectric current $I_{pyro}$ (d), and evaluated electric polarization $P$ (e), as a function of $T$ respectively, for sample $x=0.0$. The decomposed $P_{DM}$ and $P_{MM}$ are shown in (e) [??].

Fig.7. Measured magnetization $M$ for samples $x=0$, 0.04, 0.08, and 0.20, as a function of $T$ respectively.

Fig.8. Measured pyroelectric current $I_{pyro}(T)$ and evaluated polarization $P(T)$ for a series of samples in (a)~(g), where the $I_{pyro}(T)$ data for sample $x=0.0$ are inserted for reference. The $P$ data at several temperatures as a function of $x$ are plotted in (h).

Fig.9. Evaluated $P_{MM}$ and $P_{DM}$ as a function of $x$ respectively at $T=2K$.

Fig.10. The measured phase diagram on the $(x, T)$ space. The crossing between region I and region II is indicated by the solid double head arrow, suggesting the reversing of electric polarization $P$. In region III, no more polarization component $P_{MM}$ is available.

Fig.11. Measured $P(T)$ data under various magnetic field $H$ for sample $x=0.0$ (left column), $x=0.12$ (middle column), and $x=0.16$ (right column).

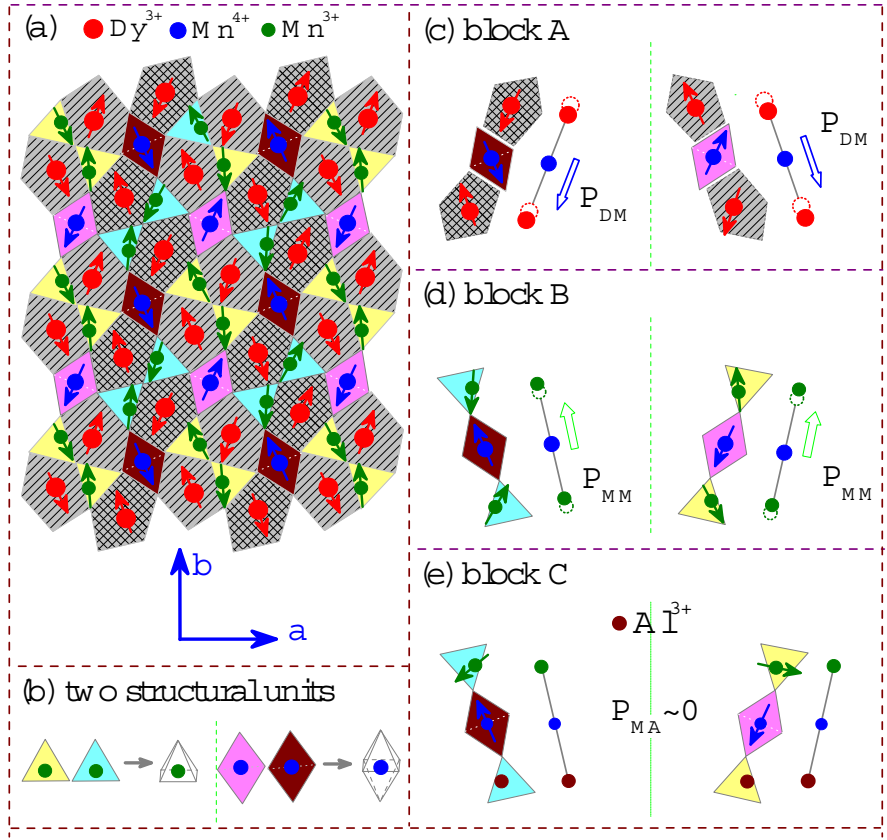

Fig.1. A schematic drawing of the spin structure at low T and possible mechanisms for electric polarization generation. (a) The spin structure projected on the *ab* plane, and these structural units are not on the same lattice plane and they alternatively shift roughly 1/2 or -1/2 atomic layer distance along the *c*-axis. (b) The structural unit symbols used in (a). (c) The two sub-types of block A, and the local electric dipoles $P_{DM}$ due to the symmetric exchange striction associated with the ↓↓↑ and ↑↑↓ $Dy^{3+}$-$Mn^{4+}$-$Dy^{3+}$ spin alignments, respectively. (d) The two sub-types of block B, and the local electric dipoles $P_{MM}$ due to the symmetric exchange striction associated with the ↓↑↑ and ↑↓↓ $Mn^{3+}$-$Mn^{4+}$-$Mn^{3+}$ spin alignments, respectively. (e) The two sub-types of block C with the Al-substituted $Mn^{3+}$-$Mn^{4+}$-$Al^{3+}$ chains which have no electric dipole. The lattice of $DyMn_2O_5$ thus accommodates a ferrielectricity.

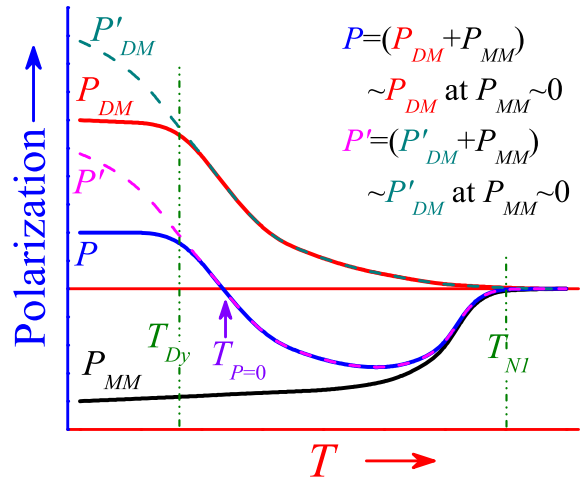

Fig.2. The predicted *T*-dependence of electric polarization *P* as a feature of ferrielectricity, where polarization *P* consists of two anti-parallel components $P_{DM}$ and $P_{MM}$. It is suggested that these polarizations should take the dependences *P′*, $P′_{DM}$ and $P_{DM}$ if no independent Dy spin ordering would occur below $T_{Dy}$.

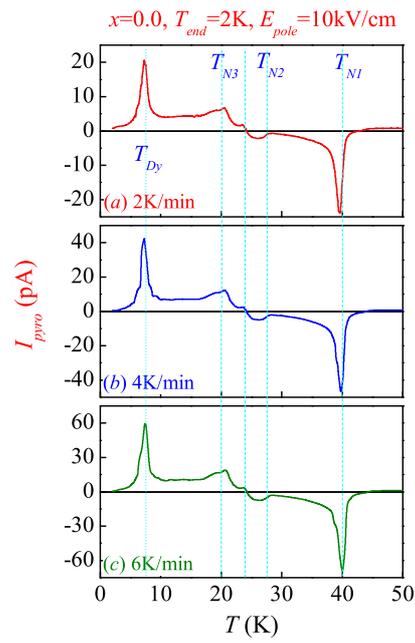

Fig.3. Measured pyroelectric current $I_{pyro}(T)$ for DyMn2O5 with warming rate of 2K/min (a), 4K/min (a), and 6K/min (a).

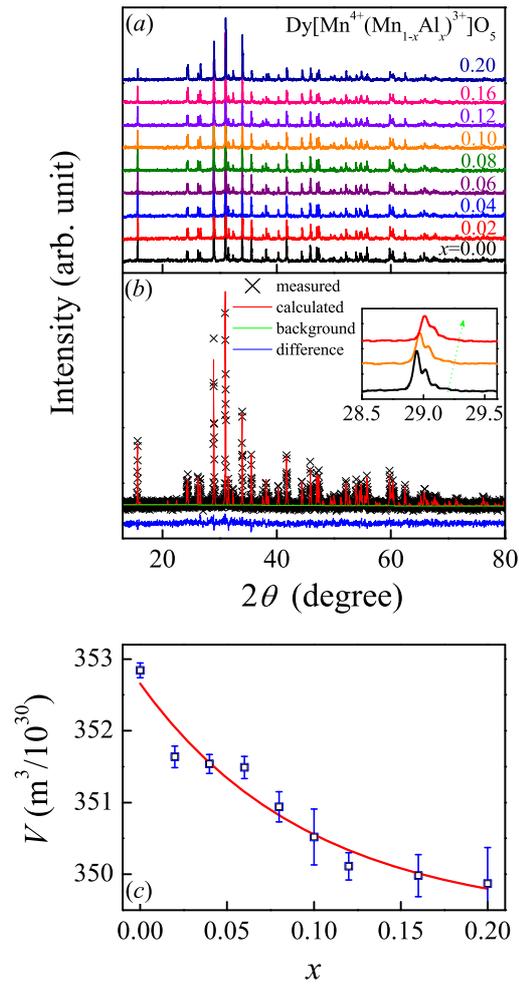

Fig.4. (a) Measured XRD $\theta$-$2\theta$ spectra for a series of DyMn$_{2-x}$Al$_x$O$_5$ samples with labeled $x$ values. (b) The Rietveld refined data for sample $x$=0.04 and the insert shows the local reflections of three samples $x$=0.00, 0.10, and 0.16. (c) The fitted unit volume V as a function of $x$ with a fitting curve for guide of eyes.

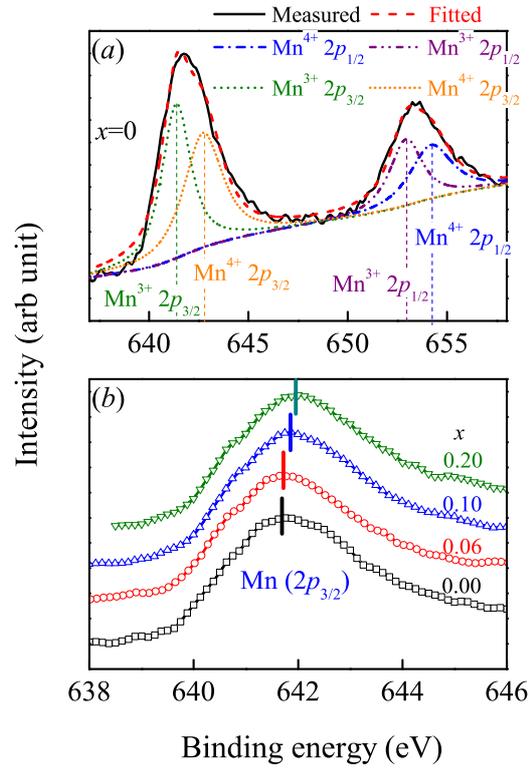

Fig.5. (a) Measured XPS spectrum for sample $x$=0.0 as an example, where the $Mn^{3+}$ and $Mn^{4+}$ core energy levels are labeled and the measured peaks are decomposed as the superimposition of the contributions from the $Mn^{3+}$ and $Mn^{4+}$ excitations, respectively. (b) The local peaks around 642 eV corresponding to the Mn ($2p_{3/2}$) core level. The peak shifting toward the high energy with increasing $x$ is shown, indicating the increasing $Mn^{4+}/Mn^{3+}$ ratio.

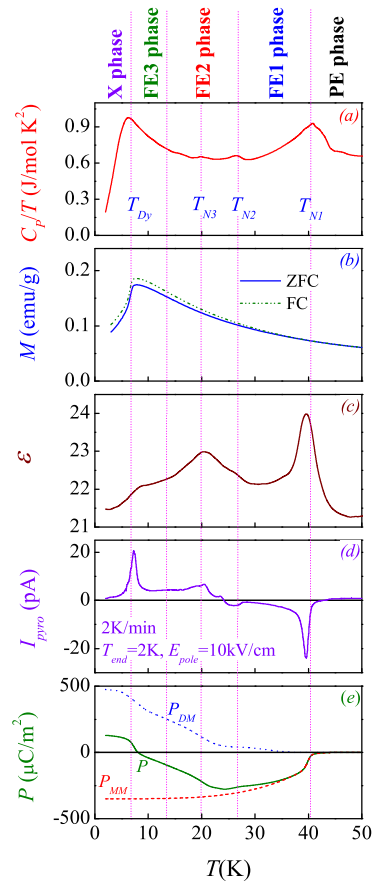

Fig.6. Measured specific heat $C_P/T$ (a), magnetization $M$ (b), dielectric constant $\varepsilon$ (c), pyroelectric current $I_{pyro}$ (d), and evaluated electric polarization $P$ (e), as a function of $T$ respectively, for sample $x$=0.0. The decomposed $P_{DM}$ and $P_{MM}$ are shown in (e) [??].

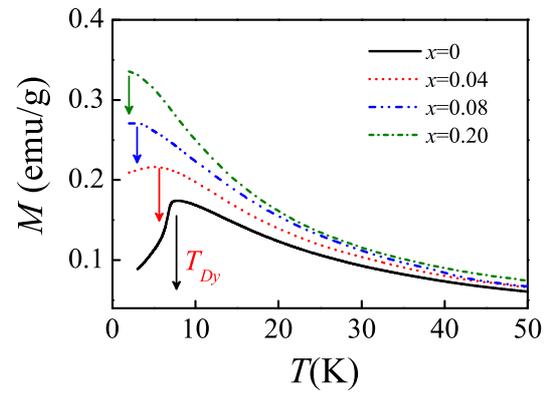

Fig.7. Measured magnetization *M* for samples *x*=0, 0.04, 0.08, and 0.20, as a function of *T* respectively.

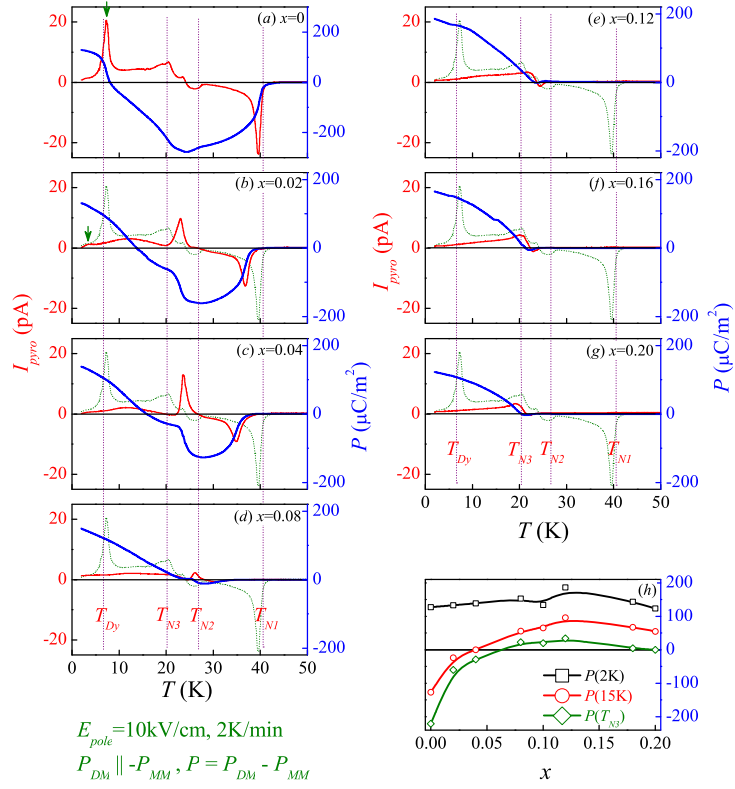

Fig.8. Measured pyroelectric current $I_{pyro}(T)$ and evaluated polarization $P(T)$ for a series of samples in (a)~(g), where the $I_{pyro}(T)$ data for sample $x$=0.0 are inserted for reference. The $P$ data at several temperatures as a function of $x$ are plotted in (h).

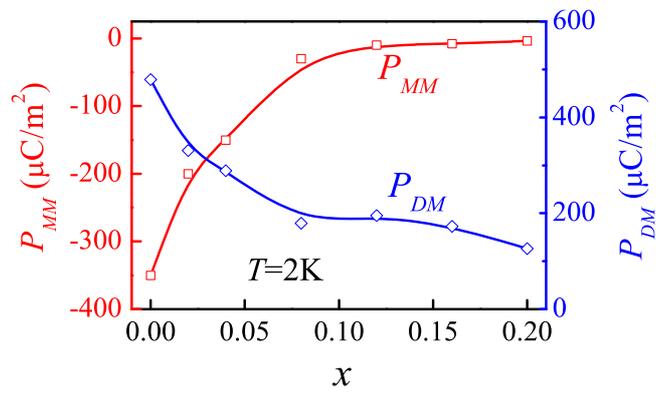

Fig.9. Evaluated $P_{MM}$ and $P_{DM}$ as a function of $x$ respectively at $T$=2K.

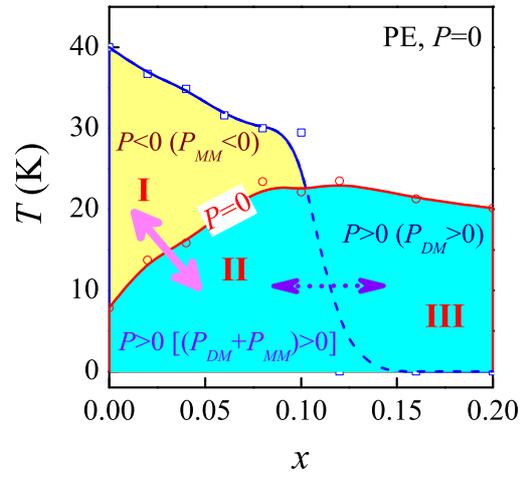

Fig.10. The measured phase diagram on the ($x$, $T$) space. The crossing between region I and region II is indicated by the solid double head arrow, suggesting the reversing of electric polarization $P$. In region III, no more polarization component $P_{MM}$ is available.

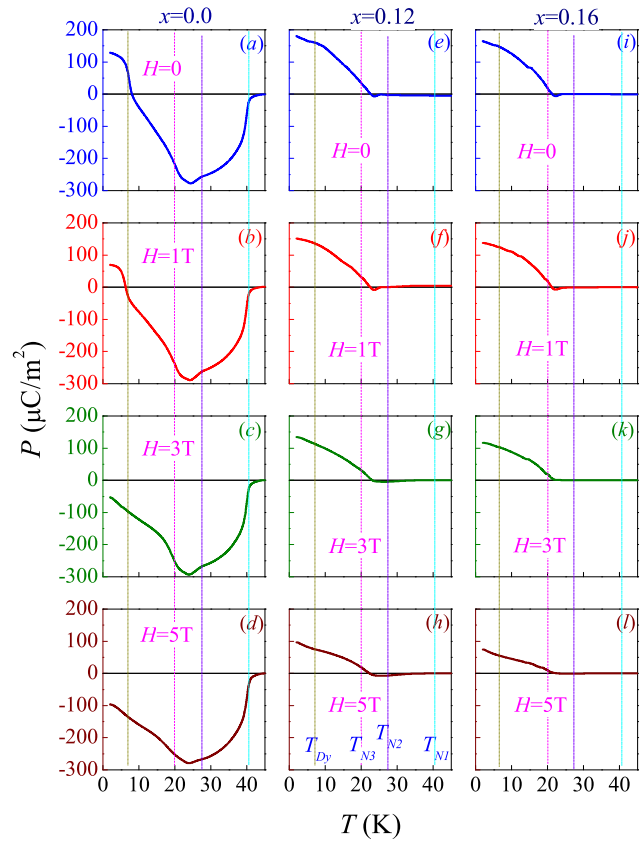

Fig.11. Measured $P(T)$ data under various magnetic field $H$ for sample $x$=0.0 (left column), $x$=0.12 (middle column), and $x$=0.16 (right column).